\documentclass[
  pra,       % Estilo de Physical Review Applied
  superscriptaddress,
  aps,             % American Physical Society (siempre presente)
  reprint,         % Formato de dos columnas (simula la versión publicada)
  floatfix,        % Evita que figuras "escapen" al final del documento
  amsmath,         % Soporte extendido de matemáticas AMS
  amssymb         % Símbolos matemáticos adicionales
]{revtex4-2}

\usepackage[utf8]{inputenc}   % Lets you type UTF-8 characters (accents, etc.) directly
\usepackage[T1]{fontenc}      % Correct font encoding
\usepackage{graphicx}       % Inserción de figuras (includegraphics)
\usepackage{dcolumn}        % Alineación decimal en tablas
\usepackage{bm}             % Símbolos en negrita en modo matemático (\bm{})
\usepackage[normalem]{ulem}
\usepackage{soul}
\usepackage{xcolor}         % Colores (útil para notas de revisión)
\usepackage{physics}        % Notación física conveniente (\bra, \ket, \abs...)
\usepackage{amsmath}

\usepackage{amssymb}
\usepackage{amsthm}
\usepackage{bm}
\usepackage{mathrsfs}
\usepackage{mathtools}
\usepackage{braket}

\usepackage{graphicx}
\usepackage[percent]{overpic}
\usepackage{tikz}
\usepackage{dcolumn}
\usepackage{booktabs}
\usepackage{multirow}

\usepackage{xcolor}
\usepackage{soul}
\usepackage{comment}

\usepackage{ragged2e}
\usepackage{microtype}

\usepackage{wasysym}

\definecolor{olivegreen}{RGB}{51,102,0}
\definecolor{DodgerBlue}{RGB}{0,90,156}
\definecolor{orangevillavics}{RGB}{204,102,0}
\definecolor{redjesus}{RGB}{204,0,102}
\definecolor{greenfernando}{RGB}{51,102,0}
\definecolor{darkred}{rgb}{0.8,0.1,0.1}
\definecolor{darkblue}{rgb}{0.1,0.1,0.7}

\begin{document}

\title{Quantum Channel-Induced Geometry of the Uhlmann Phase in Qubit Systems}

\author{F. Nieto-Guadarrama}%[orcid=0000-0003-4228-5182]
    \email{nieto.fernando@ens.cnyn.unam.mx}
    \affiliation{Centro de Nanociencias y Nanotecnolog\'ia, Universidad
    Nacional Aut\'onoma de M\'exico,
            Km 107 Carretera Tijuana-Ensenada, 
            Ensenada,
            22800, 
            B.C.,
            M\'exico}
    \affiliation{Facultad de Ciencias, Universidad Aut\'onoma de Baja
            California,
            Carretera Transpeninsular Tijuana-Ensenada No. 3917, 
            Ensenada,
            22800, 
            B.C.,
            M\'exico}
\author{F. Rojas}%[orcid=0000-0002-3893-1850]
    \email{frojas@ens.cnyn.unam.mx}
    \affiliation{Centro de Nanociencias y Nanotecnolog\'ia, Universidad
    Nacional Aut\'onoma de M\'exico,
            Km 107 Carretera Tijuana-Ensenada, 
            Ensenada,
            22800, 
            B.C.,
            M\'exico}
\author{J. Villavicencio}%[orcid=0000-0002-2523-6584]
    \email{villavics@uabc.edu.mx}
    \affiliation{Facultad de Ciencias, Universidad Aut\'onoma de Baja
            California,
            Carretera Transpeninsular Tijuana-Ensenada No. 3917, 
            Ensenada,
            22800, 
            B.C.,
            M\'exico}
\author{Jes\'us A. Maytorena}%[orcid=0000-0002-8513-108X]
    \email{jesusm@ens.cnyn.unam.mx}
    \affiliation{Centro de Nanociencias y Nanotecnolog\'ia, Universidad
    Nacional Aut\'onoma de M\'exico,
            Km 107 Carretera Tijuana-Ensenada, 
            Ensenada,
            22800, 
            B.C.,
            M\'exico}
\date{\today}
\begin{abstract}

We show that a cyclically controlled non-coherence-generating channel supplies a two-parameter family of closed Bloch trajectories, absent in the single-parameter (temperature) cycles of the thermal Uhlmann literature, whose associated Uhlmann phase develops a genuine vortex--antivortex structure on the channel-parameter torus.
A qubit prepared in a pure state, which carries no geometric phase of
its own, acquires a nontrivial Uhlmann phase, obtained here in closed
form, purely  from a closed loop in the space of the native quantum channel  parameters. 
The net topological charge of these defects is constrained to zero by
the Poincar\'e--Hopf theorem.
As the input-state
orientation approaches its critical values, the defects merge and
annihilate in pairs following a square-root coalescence law. This
defect dynamics drives local geometric transitions, a mechanism that
contrasts with single-parameter thermal Uhlmann transitions, where a global quantized
winding jumps as a bulk invariant.

\end{abstract}

\maketitle

\paragraph*{Introduction.---}
Geometric phases admit inequivalent extensions to mixed quantum states.
The Uhlmann construction generalizes the Berry phase via parallel transport of density-matrix amplitudes in the purification bundle~\cite{UhlOrg,uhlmannlmp91,BerryOrg}. This approach differs from interferometric mixed-state phases and their nonunitary extensions~\cite{sjqvist2000geometric,Carollo2003,Tong2004}.
Previous studies of Uhlmann geometry have focused on Gibbs states under cyclic parameter variations, identifying finite-temperature transitions in fermionic and symmetry-protected topological band models, and in spin-$j$ and spin-coherent or spin-squeezed systems
\cite{viyuelaprl14,PhysRevLett.113.076408,PhysRevLett.113.076407,
Viyuela_2015,PhysRevA.104.023303,Wang2025}.
Our own work belongs to this line: we have analyzed spin-$j$~\cite{morachisetal_21},
composite and entangled~\cite{villa_etal_2021}, locally
driven two-spin~\cite{villa_etal_23}, and spin-1~\cite{nieto2024uhlmann}
systems.
Experimental implementations on superconducting-qubit platforms have progressed from a single qubit to an intermediate-temperature spin-$1$ regime~\cite{viyuelanpj18,Mastandrea2026}. 
In all of these works, ours included, the environment plays a
\emph{passive} role: it fixes a degree of mixedness that stays constant
during each cycle, while the closed path is generated by varying
system-Hamiltonian parameters, and the transition is a jump in a global
quantized winding number acting as a bulk invariant~\cite{viyuelaprl14}.

In this Letter we reverse both roles. First, the system Hamiltonian remains
static and the cycle is driven entirely by the channel, so that the geometric
phase is generated by the system--environment coupling rather than by
Hamiltonian control. Second, the resulting criticality is local, carried by
point defects of unit charge, rather than a jump of a global invariant.
For a non-coherence-generating (NCG) qubit
channel~\cite{huChannelsThatNot2016a}, we derive the Uhlmann phase in
closed form using its Bloch-space connection~\cite{UhlBloch}. 
Zeros of the complex holonomy amplitude form vortex--antivortex defects on the channel-parameter torus, whose  net topological charge is forced to vanish by the Poincar\'e--Hopf theorem. 
This vanishing net charge constrains the local dynamics: defects can only nucleate and annihilate in pairs, and their coalescence follows a channel-independent square-root law as the input-state orientation approaches its critical values.
The resulting criticality is thus local, yet topologically constrained, in contrast to the bulk-invariant jump of the thermal case.

\paragraph*{Channel-Induced Generation of a Geometric Phase.---}
The evolution of a mixed state \(\rho(t)\) is modeled by a completely positive and trace-preserving (CPTP) map~\cite{nielsen11,rivas2012open} with Kraus operators \(K_\ell(t,t_0;\lambda)\) satisfying \(\sum_{\ell} K_{\ell}^\dagger K_{\ell}=\mathbb{I}\), where \(\lambda\) denotes control parameters.
We demonstrate that a cyclically controlled quantum channel can induce a nontrivial Uhlmann phase, even starting from an initial state with a vanishing phase. Specifically, we consider an NCG channel, which is a CPTP map that does not generate coherence from incoherent states with respect to a fixed reference basis
~\cite{Baumgratz2014,Streltsov2017}. We employ the NCG channel from Ref.~\cite{huChannelsThatNot2016a}, characterized by
\begin{equation}
K_0 =
\begin{pmatrix}
\cos\nu & 0\\
0 & e^{i\chi}\cos\mu
\end{pmatrix},
\qquad
K_1 =
\begin{pmatrix}
0 & \sin\mu\\
e^{i\chi}\sin\nu & 0
\end{pmatrix},
\label{eq:NCG-Kraus}
\end{equation}
where $\mu,\nu,\chi$ are externally controlled channel parameters, with
$\chi$ defined modulo $2\pi$ and $(\mu,\nu)\in[0,\pi)\times[0,\pi)$,
exhausting all distinct channels in the family.
Since the channel has Kraus rank two, each instance
admits a minimal Stinespring dilation with a single ancillary
qubit~\cite{nielsen11}. Initializing the ancilla in $|0\rangle$, a
joint unitary $U(\mu,\nu,\chi)$ satisfies
$K_0=\langle 0|U(\mu,\nu,\chi)|0\rangle,$
and
$K_1=\langle 1|U(\mu,\nu,\chi)|0\rangle$.
Tracing out the ancilla recovers Eq.~\eqref{eq:NCG-Kraus}. 
For each fixed $(\mu,\nu)$, the cycle $\chi:0\to2\pi$ thus defines a
closed loop of CPTP maps. We label this family of loops using
$\alpha=\mu-\nu$ and $\beta=\mu+\nu$.
The torus $(\alpha,\beta)\in[0,2\pi)^{2}$ is a twofold cover of the physical
domain $(\mu,\nu)\in[0,\pi)^{2}$, with deck transformation
$(\alpha,\beta)\to(\alpha+\pi,\beta+\pi)$. Under it
$s_{\parallel}\to-s_{\parallel}$ and $s_{\perp}\to s_{\perp}$; since
$s_{\parallel}$ enters the holonomy amplitude only through
$s_{\parallel}^{2}$ [Eq.~\eqref{eq:uhlmann_phase_qubit} below], the Uhlmann
phase is invariant.
We therefore work on the covering torus, where the
topological bookkeeping is cleanest, each physically distinct loop appearing
twice.
The NCG constraint isolates the role of the channel-induced trajectory:
diagonal states remain diagonal in the reference basis, while any
pre-existing coherences are only attenuated and may acquire a phase.
\begin{figure}
    \centering
    \includegraphics[width=1\linewidth]{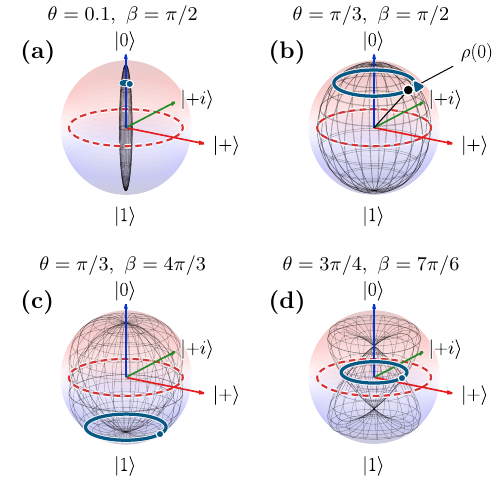}
    \caption{\justifying Bloch-vector trajectories generated by the NCG channel at the
$(\theta,\beta)$ indicated above each panel. The mesh depicts the closed surface
$\vec r(\alpha,\chi)$ of Eq.~\eqref{eq:bloch-trajectory}, swept by
$\alpha,\chi\in[0,2\pi)$: a thin spindle in (a), a ellipsoid in (b), and
folded surfaces with self-intersections in (c) and (d). In every panel the
thick dark-blue curve is the representative Uhlmann cycle
$\vec r(\pi/4,\chi)$, traversed counterclockwise when viewed from $+z$; in
(a) it is nearly point-like, the transverse radius $s_{\parallel}$ being
strongly suppressed at small $\theta$. The dot marks the reference point
$\rho(\chi=0)$ and the arrowhead the sense of increasing $\chi$, both
labeled only in (b). The red dashed circle of radius $\sqrt{3}/2$ in the
equatorial plane ($s_{\perp}=0$, $s_{\parallel}^{2}=3/4$) is where the
Uhlmann phase becomes undefined (see the text).}
    \label{fig:ncg-bloch}
\end{figure}

As input we take the ground state of a spin-$\frac{1}{2}$ particle in a
magnetic field lying in the $x$-$z$ plane, 
$H=-(\sin\theta\,\sigma_x+\cos\theta\,\sigma_z)$ with $\theta\in[0,\pi]$:
at zero temperature the system occupies
$\ket{g}=\cos(\theta/2)\ket{0}+\sin(\theta/2)\ket{1}$, so the initial
density matrix is 
$\rho_g=\ket{g}\!\bra{g}$. 
The channel output can be written as 
$\rho(\chi)=\left[\mathrm{I}+\vec r(\chi)\cdot\vec\sigma\right]/2$ with
\begin{equation}
\begin{aligned}
\vec r(\chi)
&=\bigl(s_{\parallel}\cos\chi,\,
        s_{\parallel}\sin\chi,\,
        s_{\perp}\bigr),\\
s_{\parallel}
&=\cos\alpha\,\sin\theta,\\
s_{\perp}
&=\cos\theta\cos\alpha\cos\beta+\sin\alpha\sin\beta,
\end{aligned}
\label{eq:bloch-trajectory}
\end{equation}
where  $\vec{\sigma}=(\sigma_x,\sigma_y,\sigma_z)$.
The NCG character of the channel is reflected in the bound
$\mathcal{C}=|\cos\alpha|\sin\theta \leq \sin\theta$, so that the
output state never carries more coherence than the initial pure state.

We highlight the dependence on the channel parameter $\chi$ because it is its
cyclic variation that is used to generate the Uhlmann holonomy. Note, however, that the Bloch vector
(\ref{eq:bloch-trajectory}) depends also on the channel parameters $\alpha, \beta$ and the
qubit direction $\theta$. For fixed ($\beta,\theta$), the vector (\ref{eq:bloch-trajectory})
defines a parametric surface $\vec{r}(\alpha,\chi)$ within the Bloch ball, where 
the signed $s_{\parallel}$ and $s_{\perp}$ become functions of $\alpha$ only.
Thus, for any given value of $\alpha$, varying $\chi \in [0,2\pi)$ traces a horizontal closed trajectory.
This is illustrated in Fig.~\ref{fig:ncg-bloch}, which displays some examples of such a surface and how it evolves 
for several values of the pair ($\beta,\theta$). For $\beta=\pi/2$ the surface is the ellipsoid
$s_{\parallel}^2+\sin^2\theta\,s_{\perp}^2=\sin^2\theta$, which evolves between a thin spindle
for $\theta\approx 0$ [Fig.~\ref{fig:ncg-bloch}(a)] and the unit sphere for $\theta=\pi/2$;
Fig.~\ref{fig:ncg-bloch}(b) shows the case for $\theta=\pi/3$, 
$s_{\parallel}^2+(3/4)s_{\perp}^2=3/4$. The effect of the NCG channel when $\beta\neq\pi/2$
is illustrated in Fig.~\ref{fig:ncg-bloch}(c) and (d). Now the locus folds onto itself, 
producing self-intersections. 
Each highlighted blue path is the $\chi$-cycle obtained using the
common choice $\alpha=\pi/4$, for which $s_{\perp}$ remains constant
along the cycle.
When $|\cos\alpha|=0$ the coherence $\mathcal{C}$ vanishes and the
corresponding $\chi$-cycle degenerates to a single point on the
$|0\rangle$--$|1\rangle$ axis, $(0,0,\pm\sin\beta)$; there the swept
surface closes onto the axis, producing the two pinch points visible
as the spindle tips in (a).

The Uhlmann phase is thus evaluated along cycles generated directly by the channel, whose intrinsic geometry delineates where topological defects nucleate.
For parameters such that $r=|\vec r|<1$, Eq.~\eqref{eq:bloch-trajectory}
defines a full-rank trajectory, along which the Uhlmann phase
$\Phi_U$~\cite{UhlOrg} and the Bloch-space connection
$\mathcal{A}$~\cite{UhlBloch} are
\begin{equation}
\Phi_U =
\arg\!\left[
\operatorname{Tr}\!\left(
\rho_0 \, \mathcal{P}
e^{\oint_C \mathcal{A}}
\right)
\right], \quad
\mathcal{A}
    =
    \frac{i}{2}
    \frac{\vec{\sigma}\cdot(d\vec r\times \vec r)}
    {1+\sqrt{1-r^2}}, 
\label{eq:uhlmann-general}
\end{equation}
where $\mathcal{P}$ is the path-ordering operator, $\rho_0 = \rho(0)$ is the initial density matrix at the start of the closed trajectory $C$, and $\mathcal{A}$ is the Uhlmann connection that generates the parallel transport of purifications along $C$. 
For the NCG channel defined in Eq.~\eqref{eq:NCG-Kraus}, the longitudinal component $s_{\perp}$ is independent of the parameter \(\chi\), while the transverse components undergo a rigid rotation. Consequently, the Uhlmann connection simplifies to%
\begin{equation}
    \mathcal{A} = \frac{i}{2}e^{-i\chi \sigma_z/2}\left( \frac{s_{\perp}s_{\parallel} \sigma_x- s_{\parallel}^2 \sigma_z}{1+\sqrt{1-r^2}}\right) e^{i\chi \sigma_z/2}\, d\chi. \label{A}
\end{equation}
\begin{figure*}[t]
  \centering
  \includegraphics[width=\textwidth]{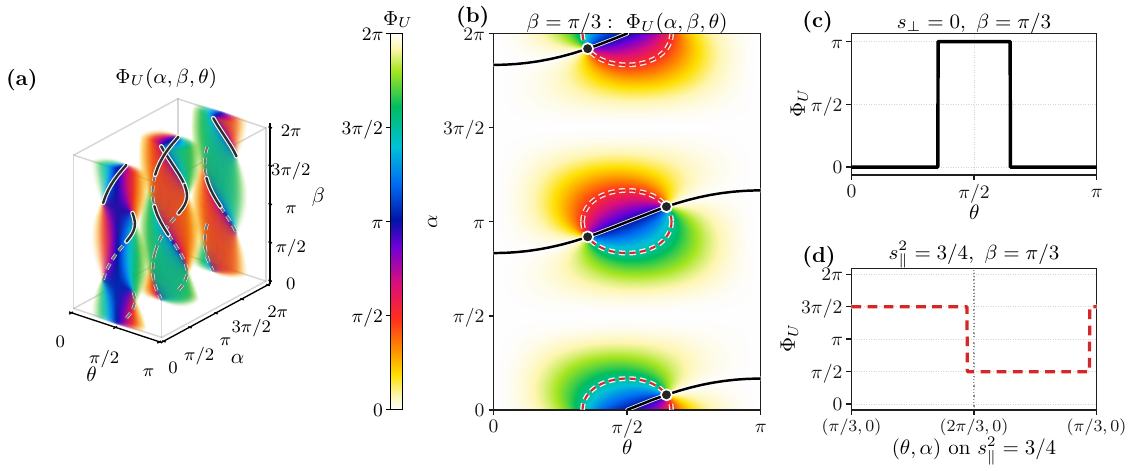}
\caption{
(a) $\Phi_U(\alpha,\beta,\theta)$ over the parameter cube; values
near $0$ and $2\pi$ are rendered transparent to reveal the interior
structure; dark neutral curves with thin white halos mark the critical
locus. (b) Slice at $\beta=\pi/3$ in the $(\theta,\alpha)$ plane:
solid black curves mark $s_{\perp}=0$, red dashed loops mark
$s_{\parallel}^2=3/4$, and their intersections, shown as dark filled circles
with thin white outlines, are the phase vortices. (c), (d) $\Phi_U$ along
$s_{\perp}=0$ and $s_{\parallel}^2=3/4$, respectively, in the same slice;
the discontinuities occur at the intersections of the two loci and thus
locate the vortices. The lower axis in (d) gives the corresponding
$(\theta,\alpha)$ coordinates along the loop.}
  \label{fig:UhlPhase}
\end{figure*}
\begin{figure*}[t]
    \centering
    \includegraphics[width=0.9\linewidth]{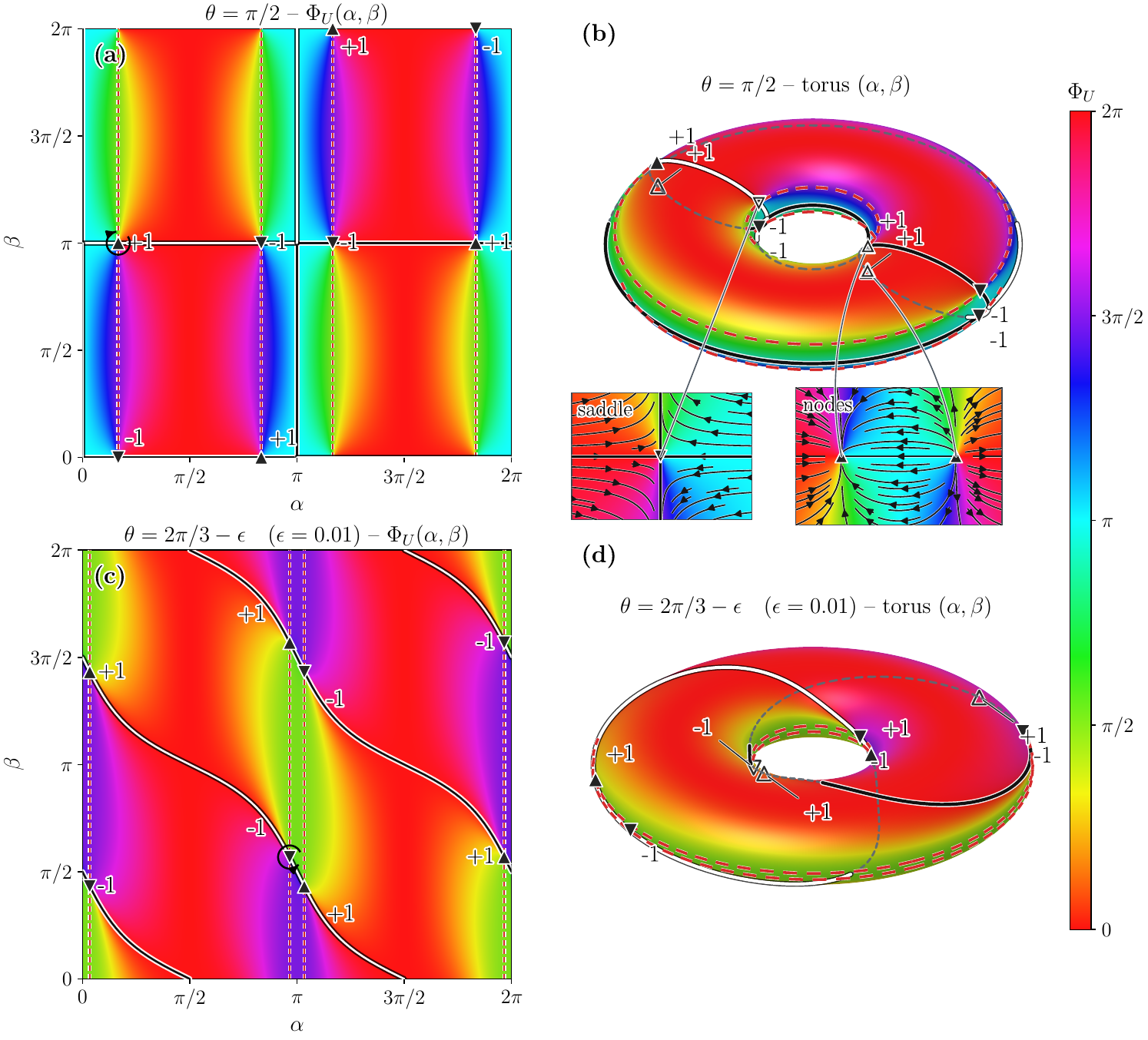}
    \caption{Uhlmann phase $\Phi_U(\alpha,\beta)$ generated by the NCG channel at
$\theta=\pi/2$ [(a), (b)] and $\theta=2\pi/3-\epsilon$ with $\epsilon=0.01$
[(c), (d)], on the fundamental square of the channel parameters [(a), (c)]
and on the covering torus $T^{2}$ [(b), (d)]. The color scale is cyclic:
$\Phi_U=0$ and $\Phi_U=2\pi$ share the same color. Solid black and white
curves are the two branches of $s_{\perp}=0$, red dashed curves are
$s_{\parallel}^{2}=3/4$, and their crossings are the vortices,
where $\Phi_U$ is undefined. Winding numbers $[+1,-1]$ are marked respectively by
[$\blacktriangle$, $\blacktriangledown$] ([$\vartriangle$, $\triangledown$]) on the visible (hidden) portion of the torus [(b), (d)]. The circular arrow gives the sense in which $\Phi_U$ circulates. The insets in (b) show $\mathcal{G}$ as a vector field:
by the Jacobian of Eq.~\eqref{eq:winding}, a $w=-1$ vortex is a saddle point, and a
$w=+1$ vortex a source or sink node.}
    \label{fig:uhlmann_torus}
\end{figure*}
\begin{figure*}[!htbp]
    \centering
    \centering
    \includegraphics[width=1\linewidth]{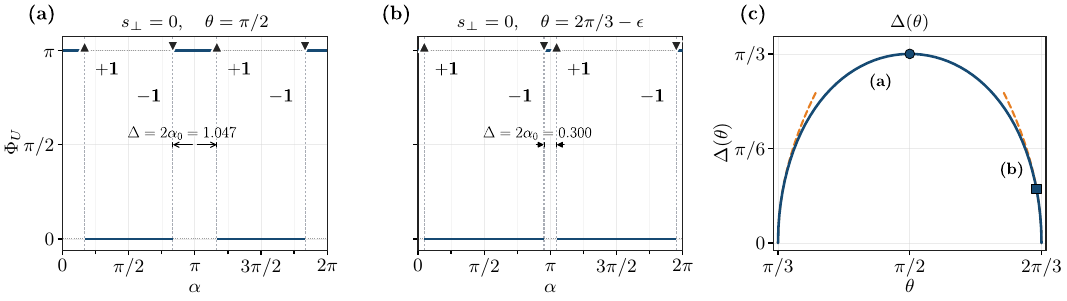}
\caption{Uhlmann phase $\Phi_U$ as a function of $\alpha$ along the locus
$s_{\perp}=0$, for (a) $\theta=\pi/2$ and (b) $\theta=2\pi/3-\epsilon$, with
$\epsilon=0.02$. The $\Phi_U=\pi$ plateau is bounded by 
$[+1,-1]$ vortices marked respectively by [$\blacktriangle$, $\blacktriangledown$];
neutral vertical dashed lines indicate their
positions. Its width equals the vortex separation
$\Delta(\theta)$, decreasing from $\Delta=\pi/3\simeq1.047$ in (a) to
$\Delta\simeq0.300$ in (b). Panel (c) shows the exact separation from
Eq.~\eqref{eq:Delta} (solid blue curve), and the near-edge square-root law
$\Delta(\theta)\propto (\theta-\theta_c)^{1/2}$
(orange dashed curves). The filled markers
identify the parameters used in panels (a) and (b), making the pairwise
vortex coalescence directly visible.}
    \label{fig:NGCrz}
\end{figure*}
For one complete cycle, \(0\leq\chi\leq2\pi\), the trace in
Eq.~\eqref{eq:uhlmann-general} reduces to a single complex number, the
cyclic holonomy amplitude, whose argument is the Uhlmann phase:
\begin{equation}
    \mathcal{G}(\alpha,\beta,\theta)
    =
    -\cos(\pi N)
    -i\,\frac{\sin(\pi N)}{N}\,s_{\perp},
    \,
    \Phi_U=\arg\mathcal{G},
    \label{eq:uhlmann_phase_qubit}
\end{equation}
where $N=(1-s_{\parallel}^{2})^{1/2}$. Substituting
$s_{\parallel}=\cos\alpha\,\sin\theta$ from
Eq.~\eqref{eq:bloch-trajectory} gives
$N^{2}=1-\cos^{2}\alpha\,\sin^{2}\theta$, so that $\mathcal{G}$, and
hence $\Phi_U$, depends on the trajectory only through the pair
$(s_{\parallel},s_{\perp})$. The Uhlmann phase is undefined precisely
where $\mathcal{G}$ vanishes, which yields the conditions for a phase singularity
(a `vortex' or `defect')
\begin{equation}
s_{\perp}=0,
\qquad
s_{\parallel}^{2}=3/4.
\label{eq:critical-conditions}
\end{equation}
In the Bloch ball, the latter conditions correspond
to a circle of radius $|s_{\parallel}|=\sqrt{3}/2$ on the equatorial plane
(shown as a red dashed circle in Fig.~\ref{fig:ncg-bloch}). 
Given that $|s_{\parallel}|\leqslant\sin\theta$, Eq.~(\ref{eq:critical-conditions})
implies $\theta\in[\pi/3,2\pi/3]$. Only the surface of Fig.~\ref{fig:ncg-bloch}(b)
contains this circle; in (c) the cycles with $|s_{\parallel}|=\sqrt{3}/2$ sit at
$|s_{\perp}|=1/4$ instead, and in (a) and (d) $\sin\theta<\sqrt{3}/2$ forbids contact.
Away from this critical circle, the constraint \(s_{\perp}=0\) ensures that 
$\mathcal{G}$
remains purely real, thereby restricting the Uhlmann phase to the discrete values of either \(0\) or \(\pi\).

Figure~\ref{fig:UhlPhase}(a) presents the resulting Uhlmann phase as a
three-dimensional colormap over the parameter space
$(\alpha,\beta,\theta)$. Dark curves mark the critical points  where
the phase is undefined.
The $\beta=\pi/3$ slice in (b) shows that the phase jumps occur precisely
at the intersections of the loci $s_{\perp}=0$ and $s_{\parallel}^2=3/4$; the
one-dimensional cuts in (c) and (d) follow each locus individually.
Along $s_{\perp}=0$, $\mathrm{Im}\,\mathcal{G}$ vanishes and $\Phi_U$ is
restricted to the discrete values $\{0,\pi\}$, with $\Phi_U=\pi$ exactly
where $s_{\parallel}^2>3/4$ (i.e., $N<1/2$) and $\Phi_U=0$ otherwise.
Along $s_{\parallel}^2=3/4$, $\mathrm{Re}\,\mathcal{G}$ vanishes and
$\Phi_U$ is restricted to $\{\pi/2,3\pi/2\}$, according to the sign of
$s_{\perp}$. 
The phase is therefore undefined only where the two loci
intersect, at the points where the step transitions in (c) and
(d) coincide.
To locate the critical points, we use the two conditions for a phase singularity. The second condition in Eq.~\eqref{eq:critical-conditions}, yields  $\cos^{2}\alpha=3/(4\sin^{2}\theta)$,
which admits real solutions only when $\theta \in [\pi/3,\,2\pi/3]$. Within this interval, the equation has four solutions in $[0, 2\pi)$, namely,  $\alpha\in\bigl\{\alpha_{0},\,\pi-\alpha_{0},\,\pi+\alpha_{0},\, 2\pi-\alpha_{0}\,\bigr\}$, with 
$\alpha_{0} \equiv \arccos\!\left[\sqrt{3}/(2\sin\theta)\right]$.
Substituting this constraint into the equatorial condition $s_{\perp}=0$
yields $\tan\beta= -\cos\theta\,\cot\alpha$, which provides two
solutions, $\beta_{0}$ and $\beta_{0}+\pi$, for each admissible $\alpha$:
$\{\beta_{0},\,\beta_{0}+\pi\}$ for
$\alpha\in\{\pi-\alpha_{0},\,2\pi-\alpha_{0}\}$, and
$\{\pi-\beta_{0},\,2\pi-\beta_{0}\}$ for
$\alpha\in\{\alpha_{0},\,\pi+\alpha_{0}\}$, with
$\beta_{0}\equiv\arctan\!\bigl[\sqrt{3}\cos\theta/
\bigl(2\sqrt{\sin^{2}\theta-3/4}\bigr)\bigr]$.
Consequently, for $\theta \in (\pi/3,\,2\pi/3)$, the Uhlmann phase has
exactly $4\times2=8$ vortices on the covering torus, representing four
physically distinct defects.

\paragraph*{Singularity structure on the parameter torus.---}
Figure~\ref{fig:uhlmann_torus} shows the Uhlmann phase $\Phi_U$ over the parameter plane $(\alpha$, $\beta)$, both on the fundamental square and on the torus $T^2$, which is its natural domain under double periodicity, 
for $\theta=\pi/2$ [Fig.~\ref{fig:uhlmann_torus}(a) and (b)] and $\theta\lesssim 2\pi/3$
[Fig.~\ref{fig:uhlmann_torus}(c) and (d)].
Equation~\eqref{eq:uhlmann_phase_qubit} defines a real two-component field
$\mathcal{G}=(\mathrm{Re}\,\mathcal{G},\mathrm{Im}\,\mathcal{G})$ on the
parameter torus, whose zeros, Eq.~\eqref{eq:critical-conditions}, are the
crossings of its two nodal sets (Fig.~\ref{fig:uhlmann_torus}). Because $N$
involves the channel parameters only through $\alpha$, the set
$\mathrm{Re}\,\mathcal{G}=0$ is independent of $\beta$: it is the circles
$N=1/2$ (red dashed line). The set $\mathrm{Im}\,\mathcal{G}=0$ is $s_{\perp}=0$,
whose two branches are interchanged by the half-turn $\beta\to\beta+\pi$ and
are drawn black or white according to the sign of
$\partial_{\beta}s_{\perp}$. Their eight crossings are the eight vortices
counted above.
The topological charge of a crossing $p=(\alpha_{*},\beta_{*})$, \textit{i.e.}, the winding
of $\Phi_U$ along a small counterclockwise circuit around $p$, is the sign
of the Jacobian of $\mathcal{G}$ at $p$~\cite{Needham1997}.
Its second
term vanishes identically, and at $p$, where $N=1/2$ and $s_{\perp}=0$, one
has $\partial_{\alpha}\mathrm{Re}\,\mathcal{G}
=\pi\sin2\alpha_{*}\,\sin^{2}\theta$ and
$\partial_{\beta}\mathrm{Im}\,\mathcal{G}=-2\sin\alpha_{*}/\cos\beta_{*}$,
so that
\begin{equation}
\begin{split}
w &=\mathrm{sgn}\bigl[
\partial_{\alpha}\mathrm{Re}\,\mathcal{G}\;
\partial_{\beta}\mathrm{Im}\,\mathcal{G}
-\partial_{\beta}\mathrm{Re}\,\mathcal{G}\;
\partial_{\alpha}\mathrm{Im}\,\mathcal{G}\bigr]_{p}\\[2pt]
&=-\mathrm{sgn}\!\left(\cos\alpha_{*}/\cos\beta_{*}\right)=\pm1 .
\end{split}
\label{eq:winding}
\end{equation}
The nature of each singular point is illustrated in the insets of Fig.~\ref{fig:uhlmann_torus}(b)
through the streamlines of the vector field $\mathcal{G}$.
Neither cosine vanishes anywhere in the open window $\theta\in(\pi/3,2\pi/3)$,
so every zero is nondegenerate and carries unit charge.
Since $\beta_*\!\to\!\beta_*+\pi$ maps one branch of $s_{\perp}=0$
onto the other and reverses the sign in Eq.~\eqref{eq:winding}, the
paired zeros carry opposite charges: four of each sign on the
covering torus, and zero net winding on $T^2$, as required by the
Poincar\'e--Hopf theorem. 
The deck transformation
$(\alpha,\beta)\to(\alpha+\pi,\beta+\pi)$ instead flips both cosines
and preserves $w$, so each physical vortex is represented by two
zeros of equal charge: two vortices of each sign in the physical
domain, with vanishing net charge there as well.
Vortices exist only for $\sin^2\theta\ge 3/4$, i.e., $\theta\in[\pi/3,2\pi/3]$; outside this window the torus is defect-free. At $\theta=\pi/2$ they are maximally separated [Figs.~\ref{fig:uhlmann_torus}(a), (b)], whereas for $\theta=2\pi/3-\epsilon$ [Figs.~\ref{fig:uhlmann_torus}(c), (d)] the curve $s_{\parallel}^2=3/4$ approaches tangency with the $s_{\perp}=0$ branches and oppositely charged vortices migrate toward pairwise coalescence. As $\theta$ reaches either boundary, $\theta\to\pi/3^{+}$ or $\theta\to2\pi/3^{-}$, each $+1$ vortex annihilates with a $-1$ vortex, so the net winding stays zero throughout.
The separation between a merging pair in $\alpha$ is
\begin{equation}
    \Delta(\theta) \;=\; 2\,\alpha_{0} \;=\;
    2\arccos\!\left[\sqrt{3}/(2\sin\theta)\right],
    \label{eq:Delta}
\end{equation}
with $\alpha_{0}$ the smallest positive root of $s_{\parallel}^2=3/4$. It depends only on
$\sin\theta$ and vanishes at both edges of the admissibility interval. 
Expanding (\ref{eq:Delta}) near $\theta_{c}=\pi/3$, a  square-root coalescence law 
$\Delta(\theta)\propto (\theta-\theta_c)^{1/2}$ is found, the signature of
vortex--antivortex annihilation, with the same exponent $1/2$ that the
Ginzburg-Landau mean-field
theory gives for the order parameter at a continuous phase
transition~\cite{HOHENBERG20151};  the same result is obtained for
$\theta\to2\pi/3^{-}$.

The pairwise approach and annihilation are seen directly in
Fig.~\ref{fig:NGCrz}, which shows $\Phi_U$ along $s_{\perp}=0$ versus
$\alpha$ for $\theta=\pi/2$ (a) and $\theta=2\pi/3-\epsilon$ (b).
On this locus $\Phi_U$ is pinned to $\pi$ for $N<1/2$ and to $0$
otherwise [cf.\ Fig.~\ref{fig:UhlPhase}(c)]; the switching points are 
the $+1$ and $-1$ vortices, shown respectively as upward- and
downward-pointing dark triangles; neutral
vertical dashed lines indicate their positions
separated by the plateau width
$\Delta$ of Eq.~\eqref{eq:Delta}.
The shrinking $\Phi_U=\pi$ sector is thus the direct geometric signature
of vortex--antivortex annihilation.

\paragraph*{Summary and conclusions.---}
We have derived closed-form expressions for the Uhlmann phase of a qubit
following closed trajectories in the Bloch ball, and applied them to
a cyclically controlled NCG channel.
The central result is that the phase is generated entirely by the
environment: the initial pure state carries no geometric phase, and
it is the closed loop of the channel parameter $\chi$ that imprints
a nontrivial Uhlmann holonomy. When the channel is switched off, the
phase and its entire vortex structure disappear. The geometric phase is thus a property of the channel imprinted on the system, and the parameter torus $(\alpha,\beta)$  is its natural arena.

The criticality of this channel-induced geometry exists only
at the level of local defect charges. Each vortex carries a protected
integer winding and is created or destroyed solely in
vortex--antivortex pairs, whose coalescence obeys the square-root law,
set by the tangency of
the two critical loci rather than by any microscopic feature of the
channel. Globally, the Poincar\'e--Hopf constraint keeps the total
winding of the holonomy map at zero on both sides of the transition:
the environment reshapes the local geometry while every bulk
invariant remains untouched, in contrast with thermal Uhlmann
transitions, where a global winding jumps between quantized
values~\cite{viyuelaprl14}. 
Photonic simulators already realize arbitrary single-qubit CPTP maps~\cite{PhysRevA.102.062601}, and
extending the singularity analysis to that general class is a natural next step.

These results place the channel-induced phase singularities identified
here within reach of gate-based platforms: because the NCG map has Kraus rank
two, the joint unitary that realizes it needs a single ancilla qubit, with $\chi$  cycled
over the control period. A protocol implementing Uhlmann parallel transport and
reading out the resulting phase is left for future work.

\begin{acknowledgments}
FNG acknowledges support from SECIHTI (M\'exico).
\end{acknowledgments}

\providecommand{\noopsort}[1]{}\providecommand{\singleletter}[1]{#1}%

\end{document}